# Bayesian inversion of GPR waveforms for sub-surface material characterization: an uncertainty-aware retrieval of soil moisture and overlaying biomass properties

Ishfaq Aziz[1], Elahe Soltanaghai[2], Adam Watts[3], Mohamad Alipour[1]

[1]Civil and Environmental Engineering, University of Illinois Urbana Champaign,
[2]Computer Science, University of Illinois Urbana Champaign,
[3]Pacific Wildland Fire Sciences Laboratory, United States Forest Service

## Abstract

Accurate estimation of sub-surface properties such as moisture content and depth of soil and vegetation layers is crucial for applications spanning sub-surface condition monitoring, precision agriculture, and effective wildfire risk assessment. Soil in nature is often covered by overlaying vegetation and surface organic material, making its characterization challenging. In addition, the estimation of the properties of the overlaying layer is crucial for applications like wildfire risk assessment. This study thus proposes a Bayesian model-updating-based approach for ground penetrating radar (GPR) waveform inversion to predict moisture contents and depths of soil and overlaying material layer. Due to its high correlation with moisture contents, the dielectric permittivity of both layers were predicted with the proposed method, along with other parameters, including depth and electrical conductivity of layers. The proposed Bayesian model updating approach yields probabilistic estimates of these parameters that can provide information about the confidence and uncertainty related to the estimates. The methodology was evaluated for a diverse range of experimental data collected through laboratory and field investigations. Laboratory investigations included variations in soil moisture values, depth of the overlaying surface layer, and coarseness of its material. The field investigation included measurement of field soil moisture for sixteen days. The results demonstrated predictions consistent with time-domain reflectometry (TDR) measurements and conventional gravimetric tests. The depth of the surface layer could also be predicted with reasonable accuracy. The proposed method provides a promising approach for uncertainty-aware sub-surface parameter estimation that can enable decision-making for risk assessment across a wide range of applications.

## 1. Introduction

Determining sub-surface properties and soil water content is pivotal in a multitude of domains like efficient agriculture (Liu et al., 2016), infrastructure planning and condition monitoring (Kalogeropoulos, 2011; Kaplanvural, 2023), water resource management (Srivastava, 2017), and environmental conservation. It helps optimize irrigation, prevent droughts and floods, protect ecosystems, and inform sustainable land use planning. In wildfire management, soil moisture measurement aids in identifying wildfire hotspots (Ambadan et al., 2020) and assessing the likelihood and size of wildfires (Jensen et al., 2018; Krueger et al., 2015). It also impacts microorganism survival during wildfires and prescribed burning, which affects plant growth (Dunn et al., 1985). Hence, accurate and efficient determination of soil moisture is essential in diverse applications. Soil layers are often covered on top with layers of live or dead vegetation, dry leaves, branches, and decomposed organic materials. This makes the direct measurement of soil moisture through these surface layers challenging. Furthermore, in forest fire risk assessment and post-fire effects, rapid assessment of the depth and moisture of the surface woody layer is crucial (Rao et al., 2020). This layer acts as fuel for fires; its variable depth and moisture content can directly affect the spread and aftereffects of fires on the ecosystem (Naderpour et al., 2021; García et al., 2020). Surface fuel measurements are primary inputs to fuel mapping systems (Alipour et al., 2023; Cova et al., 2023; Shaik et al., 2023; Prichard et al., 2019; Stavros et al., 2018; Falkowski et al., 2005). Moreover, inaccurate fuel characterization results in significant errors in predicting fire perimeter and burned area (Mutlu et al., 2008), limiting operational decision-making in wildland fire response and pre-fire treatments. Hence, measuring the properties of the surface layer in addition to the underlying soil is of immense importance.

Current practices for determining sub-surface moisture typically use gravimetric analysis, which involves labor-intensive and time-consuming collection and weighing of soil samples before and after oven-drying or by using different soil sensors that measure properties such as resistance, capacitance, and electromagnetic (EM) wave propagation speed. Examples of such sensors are resistive and capacitive sensors and time-domain sensors, to name a few. Resistive soil sensors are less reliable due to degradation and are not recommended for farms, while capacitive sensors are preferred for their durability and accuracy in agricultural contexts (Josephson et al., 2021). Time-domain sensors, like Time Domain Reflectometry (TDR) sensors, have been used for moisture sensing of soil and organic material (Hanes et al., 2023; Bourgeau-Chavez et al., 2010). They offer high accuracy but are relatively costly and hard to deploy, power, and maintain in bare soil and soil with overlaying material layers.

To address these challenges, we investigate the use of ground penetrating radar (GPR) to simultaneously evaluate soil and organic surface layer moisture while also estimating the depth of the organic material accumulation. An illustration of such a layer configuration of this study is shown in Figure 1. As shown in Figure 1, a GPR antenna sends an EM wave through a transmitter ($T_x$) and receives a reflected wave through a receiver ($R_x$). The received waveform is a one-dimensional signal called an A-scan (Figure 1b), which can also be viewed by transforming it into an amplitude envelope (Figure 1c). These scan data from radar surveys are used for non-destructive evaluation (He et al., 2023), sub-surface material inspection (Wickramanayake et al., 2022), under canopy biomass sensing (Sinchi et al., 2023); surface water sensing (Serbin et al., 2005), mapping buried glacier ice (Brandt et al., 2007) and many other applications by manual interpretation or by automated and semi-automated implementations such as machine learning, surrogate modeling, and full-waveform inversion (FWI). FWI is a widely used computational technique that utilizes complete waveform information, including amplitude, phase, and frequency content of a signal, to provide detailed insights into the subsurface properties, such as dielectric permittivity (ratio of a material's permittivity, $\varepsilon$ to that of free space, $\varepsilon_o$), acoustic impedance, electrical conductivity, or material depth. Researchers have used FWI of radar data for estimating the hydraulic properties of soil (Yu et al., 2022), soil moisture mapping (Wu et al., 2019; Wu et al., 2022), and various other applications (Klewe et al., 2021; Feng et al., 2019; Xie et al., 2022; Feng et al., 2023; Haruzi et al., 2022). Waveform inversion for moisture mapping and other applications often employs deterministic optimization methods such as particle swarm optimization (Dai et al., 2021; Qin et al., 2020; Kaplanvural et al., 2020) or genetic algorithm (Godio, 2016). Deterministic inversion processes seek to find the optimal solution within a parameter space, typically based on the minimization or maximization of an objective function. However, the optimal solutions provided by deterministic inversions may not always be the most appropriate representation of complex multimodal problems or real-world data that contain noise. Further, they do not provide an indication of uncertainty and other sub-optimal solutions to the inverse problems. Uncertainty-aware estimation is necessary for probabilistic risk assessments involving soil moisture and accumulated surface material. For instance, in wildfire risk assessment, researchers examine the uncertainty in wildfire modeling, offering insights for fire managers and emphasizing key dimensions of uncertainty (Riley & Thompson, 2016). Although probabilistic soil moisture retrieval has been reported using satellite data (Arellana et al., 2023), it has primarily focused on the analysis of backscattering from bare soils without overlaying layers.

The primary ecological attributes of interest in many applications include the moisture content in soil and the biomass layer, as well as the amount of overlaying biomass. Soil and woody moisture are often estimated based on the dielectric or relative permittivity (often referred to as only 'permittivity' in this paper) of the material. Relative permittivity ($\varepsilon_r$) is highly correlated to moisture in various materials, including soil (Calabia et al., 2020; Salam et al., 2019), forest litter (Razafindratsima et al., 2017; Mai et al., 2015), concrete (Kalogeropoulos, 2011; Kaplanvural, 2023; Dinh et al., 2021), and pavements (Cao & Qadi, 2022). Hence, the soil moisture content can be obtained from estimated permittivity using empirical models (Topp et al., 1980; Dobson et al., 1985; Wang & Schmugget, 1980). The moisture content of biomass can be similarly inferred from the estimated permittivity using material-specific empirical models. The amount of biomass can also be estimated as biomass or fuel load per unit area as the product of biomass depth and its bulk density.

The present study proposes an uncertainty-aware probabilistic GPR waveform inversion approach for automated estimation of soil permittivity and moisture, as well as the permittivity and depth of the overlaying surface layer (Figure 1d). To the best of our knowledge, this study is the first to propose a method for automated soil moisture estimation in the presence of surface layers. We propose the application of Bayesian inference with waveform inversion to yield probabilistic estimates of the subsurface unknowns. The posterior probability distribution of the estimated parameters will provide information about the probability of possible sets of solutions to the inverse problem and the uncertainty related to those solutions. Furthermore, the probability distributions and uncertainties can also be leveraged to identify relatively inconsistent and less reliable predictions. The results of this study allow for the rapid assessment of soil and organic surface layers by providing probabilistic estimates that could be used in uncertainty-aware analyses and risk assessments. In upcoming sections, this paper will first explain the methodology of the proposed approach, followed by a description of the experimental setup. The methodology will then be validated through experimental investigations, which will be followed by results and relevant discussions.

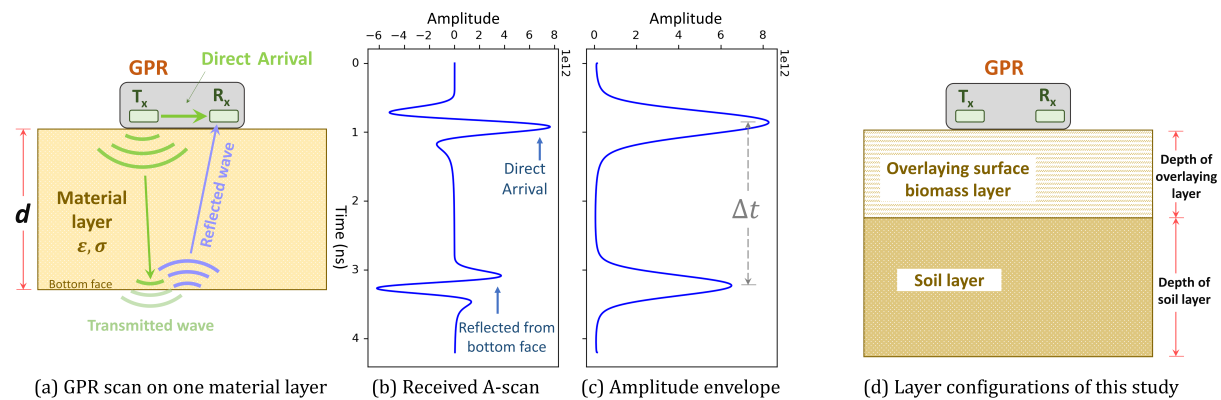

(a) GPR scan on one material layer    (b) Received A-scan    (c) Amplitude envelope    (d) Layer configurations of this study

Figure 1: (a-c) Schematic of GPR measurement and received waveform for one-layer material; (d) configuration of soil with overlaying biomass layer used in this study.

## 2. Methodology

To inspect how a GPR signal changes depending on sub-surface properties such as depth, permittivity, and electrical conductivity, assume a transmitter ($Tx$) sends an EM wave, part of which is transmitted through, and the remaining is reflected when it encounters a change in the material property (e.g., the bottom face in Figure 1a). The reflected wave is received by the radar receiver ($Rx$). The total time of travel ($\Delta t$) of the reflected wave from the transmitter to the receiver, represented by the time difference between the peaks in Figure 1(c), depends on the depth ($d$) and relative permittivity ($\varepsilon_r$). This can be utilized to analytically calculate the relative permittivity of the material layer (He et al., 2023; Lai et al., 2009), as shown in Equation 1, where $c$ is the velocity of light ($3 \times 10^8$ m/s).

$$\varepsilon_r = \left(\frac{c \times \Delta t}{2d}\right)^2 \qquad \text{Equation 1}$$

The changes in material layer properties cause changes in the received signal, as shown in Figure 2. These signals shown in this figure correspond to the one-layer configuration shown in Figure 1(a). When only depth increases, the reflected signal peak decreases in amplitude due to attenuation and shifts to the right due to increased travel time ($\Delta t$) (Figure 2a). When permittivity increases, the amplitude of direct arrival decreases, and the reflected signal peak shifts to the right (Figure 2b). If conductivity of the layer increases, the reflected wave amplitude decreases (Figure 2c).

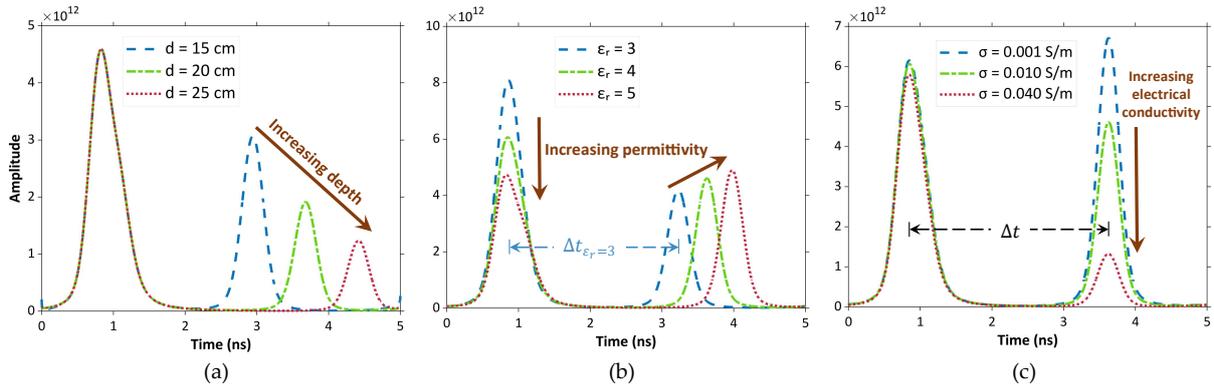

Figure 2: Variation of received GPR signal due to changes in layer properties: (a) changing layer depth, (b) changing relative permittivity, (c) changing electrical conductivity.

In this paper, we propose a methodology to leverage these signal changes to estimate subsurface parameters while accounting for the uncertainties in these estimations. Figure 3 shows the flowchart of the proposed. The first step entails the collection of the input radar signal from a GPR scan. Next, a numerical model is built using the Finite Difference Time Domain (FDTD) technique (Mescia et al., 2022). The initial material parameters of the numerical simulations are random values within a defined range. The model will then be updated by changing the parameters to match the experimental signal. Model updating can be done by an optimization algorithm, which finds the parameter values by matching the experimental signal in the numerical simulation.

Bayesian Model Updating (BMU) (Beck & Katafygiotis, 1998) was employed to obtain the probabilistic estimate of the optimal values of the parameters of interest. After the BMU analysis, Markov Chain Monte Carlo (MCMC) (Lye et al., 2021) simulation was used to obtain the posterior distribution of the parameters. At the end of the model-updating process, the numerical response obtained by using the most probable values or the mode of the probabilistic estimates more closely matches the experimental GPR signal. The estimated permittivity values can be converted to moisture contents using empirical equations like Topp's equation (Topp et al., 1980).

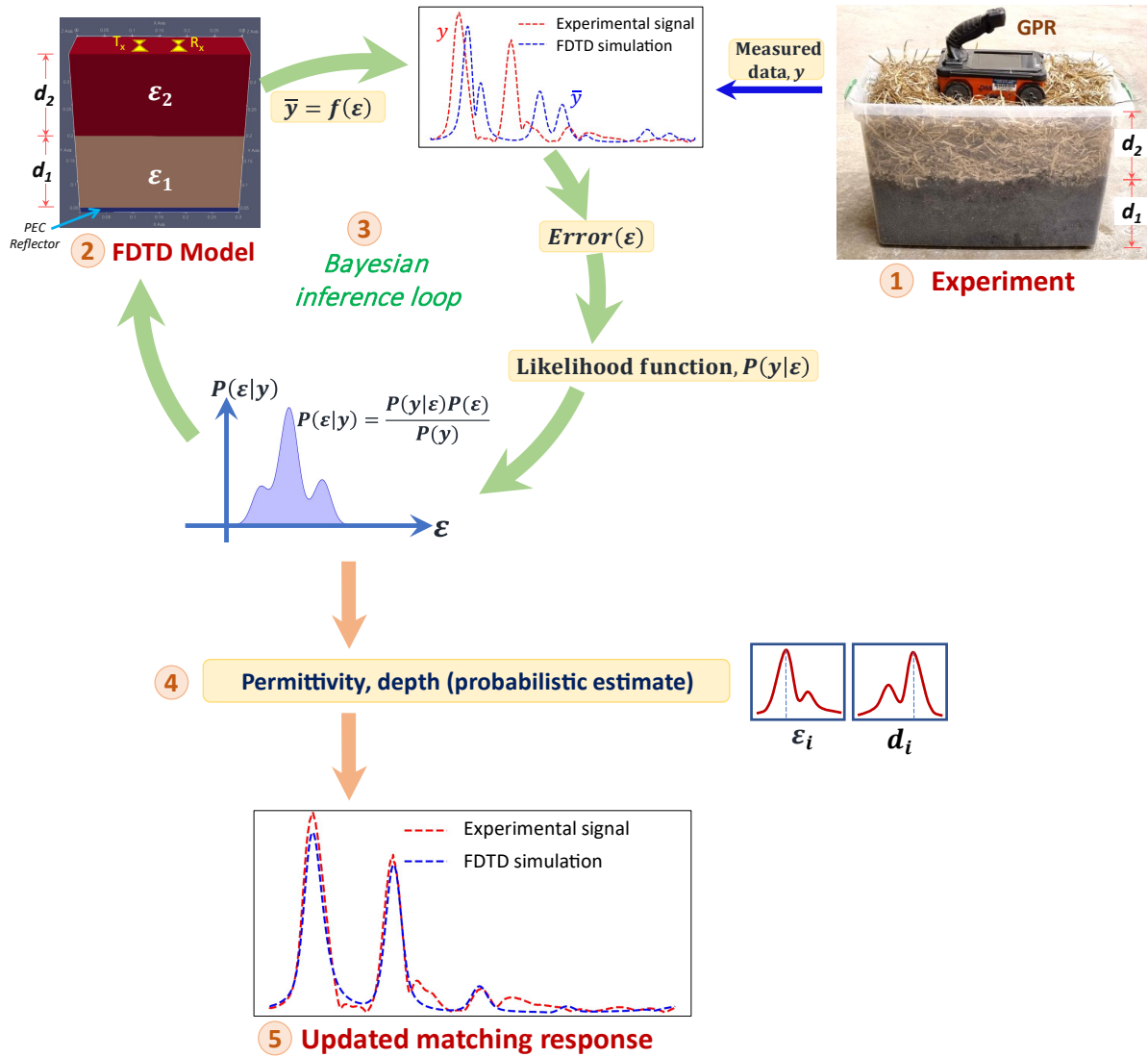

Figure 3: Proposed methodology for parameter estimation.

## 2.1. FDTD modeling

The finite difference time domain (FDTD) method is useful for solving electromagnetics problems that involve time-dependent phenomena (Mescia et al., 2022; Li et al., 2022). FDTD involves numerically solving Maxwell's equations (set of equations in Equation 2).

$$\nabla \times E = -\frac{\partial B}{\partial t}$$
$$\nabla \times H = \frac{\partial D}{\partial t} + J_C + J_s$$
$$\nabla \cdot B = 0$$
$$\nabla \cdot D = q_v$$

Equation 2

Where *E*, *B*, and *H* denote the electric field, magnetic field, and magnetic field strength, respectively. The terms *D*, $J_c$, $J_s$, $q_v$ and *t* denote electric displacement, conduction current density, displacement current density, volume electric charge density, and time (seconds), respectively. The electric field intensity at the receiver in an FDTD simulation is equivalent to the signal received by a real GPR antenna.

The FDTD method solves Maxwell's equations by discretizing both time and space. Adequate spatial discretization step sizes (Δx, Δy, and Δz) and the temporal step (Δt) are all significant in making the FDTD model more accurate and closer to the true representation. In this work, GPR wave propagation simulations were conducted using an FDTD solver called gprMax (Warren et al., 2016), which has been widely used by many researchers in the past (Liu et al., 2023; Zhang et al., 2022; Qin et al., 2021). Appropriate discretization steps have been chosen after sensitivity analysis to balance the required accuracy with computation time and avoid numerical dispersion. Figure 3 (step-2) shows an FDTD model of two material layers with a radar transmitter and receiver on the top. Perfectly matched layers (PML) around the geometry (not shown here for clarity) work as absorbent layers that absorb EM waves and represent the characteristics of an open boundary problem.

For model updating, a large number of simulations need to be performed, demanding high computation time. However, gprMax can accelerate this process by executing simulations in graphics processing units (GPUs) using NVIDIA's CUDA framework (Warren et al., 2019).

## 2.2. Antenna optimization for simulating real GPR source wavelet

An ultra-wideband (UWB) GPR manufactured by GSSI with a listed center frequency of 2700 MHz was used to conduct the experimental investigations (Geophysical Survey Systems, n.d.). This frequency was selected based on the focus of this work on surface moisture in the top 15 cm of soil and the ability of high-frequency GPR to determine shallow soil water content (Zhou et al., 2019).

An FDTD model was created to accurately simulate the transmitted wavelet of this GPR source. In most GPR simulations, theoretical sources such as the Hertzian dipole are used as transmitting antennae (Warren & Giannopoulos, 2009). However, the responses obtained from numerical simulation using such theoretical sources do not closely match the signals acquired from real GPR antennas. This is because the simplified theoretical sources do not account for the physical structure and dielectric properties of a real antenna. Some properties of a real antenna are not known or cannot be easily measured (Giannakis et al., 2018). Thus, it is customary to model and calibrate the transmitted pulse so that the simulation response closely matches the experimental data (Stadler & Igel, 2022; Warren & Giannopoulos, 2011). Although comprehensive 3D modeling of the antenna structure in FDTD simulation can be conducted for simulating real GPR response, this process requires information not readily supplied by manufacturers. On the contrary, optimizing only the transmitting pulse and its properties would be more convenient and computationally efficient. Therefore, the parameters of the transmitted pulse were optimized prior to the numerical modeling of real GPR signals. The frequency band of the GPR used in this study is 0.5-5.0 GHz, within which the center frequency lies. Although the GPR is listed to have a 2.7 GHz antenna, analysis of simple data collection scenarios showed significant deviation from this center frequency. Hence, the center frequency was used as a variable parameter. The second parameter was the type of waveform transmitted by the antenna. The waveforms used in optimization are Gaussian, Gaussiandot, Gaussiandotnorm, and Ricker. The bistatic separation (distance between the transmitter and receiver) was fixed at 6 cm as per the manufacturer's specifications. The optimum waveform and frequency were found by optimization to match the real GPR response.

To perform the antenna optimization, GPR signals were collected in the air for different distances ($d$) between the GPR and a metallic reflector plate, as shown in the experimental setup scheme of Figure 4. Two signals for distances of 24 cm and 43 cm were used as experimental responses, and the same setups were modeled in the FDTD simulation to record the numerical response. The metallic reflector was modeled in FDTD simulation as a perfect electrical conductor (PEC). Responses from both experiments and simulations were transformed into amplitude envelopes by applying the Hilbert transform. Bayesian optimization was employed to match the numerical responses to the experimental responses using the objective function (relative error) presented in Equation 3, where $y$ and $\bar{y}$ are experimental and numerical responses, respectively. For the case of the reflector at 24 cm from the GPR, the overlaid responses (amplitude envelopes) before and after optimization are shown in Figure 4. The optimum center frequency was found to be 1.579 GHz, and the optimum waveform type was Gaussian.

$$RE = \sqrt{\frac{\sum(y - \bar{y})^2}{\sum y^2}} \times 100\% \qquad \text{Equation 3}$$

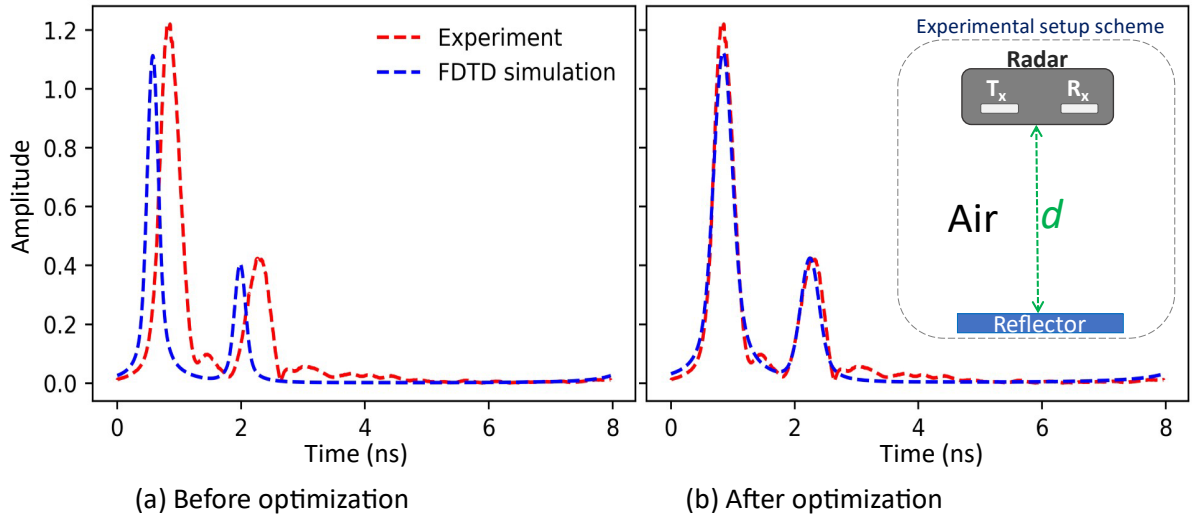

(a) Before optimization    (b) After optimization

Figure 4: Numerical and experimental response of GPR after antenna optimization.

The performance of the optimized pulse in the simulation was further validated for experimental data collected with a metal plate at varying distances from the GPR. Corresponding numerical models were created with the optimized pulse parameters, the metal plate was modeled as a PEC, and the distance of the PEC from the GPR was predicted by model updating. The predicted distances (or depths) of the metal plate with the optimized antenna pulse parameters are shown against true distances in Figure 5, which depicts reasonable accuracy. Thus, the optimized pulse parameters were used for further FDTD simulations to model the real GPR used in this study.

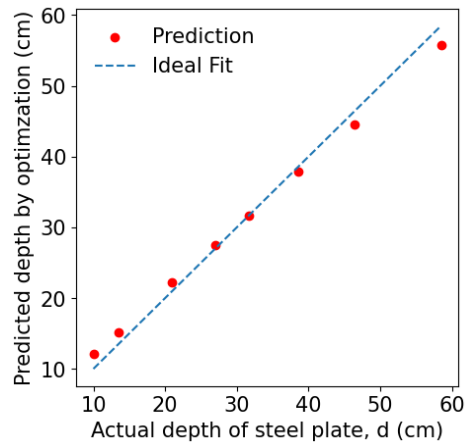

Figure 5: Validation results of antenna pulse optimization.

## 2.3. Parameter estimation

The process of estimating the parameters includes model updating to match the observed data from the experiments. For instance, in Figure 6, the signals from the numerical simulation initially do not match the experimental signal. The model is updated iteratively by changing its parameters, such as depth and relative permittivity, until the two signals match reasonably.

This results in the estimation of the optimal model and parameters that represent real-world experimental signals with reasonable accuracy.

Model updating-based waveform inversion for parameter estimation is presented using the following two approaches, viz., Bayesian optimization and Bayesian model updating with the MCMC algorithm. Bayesian model updating or Bayesian inference are synonymously used in this paper.

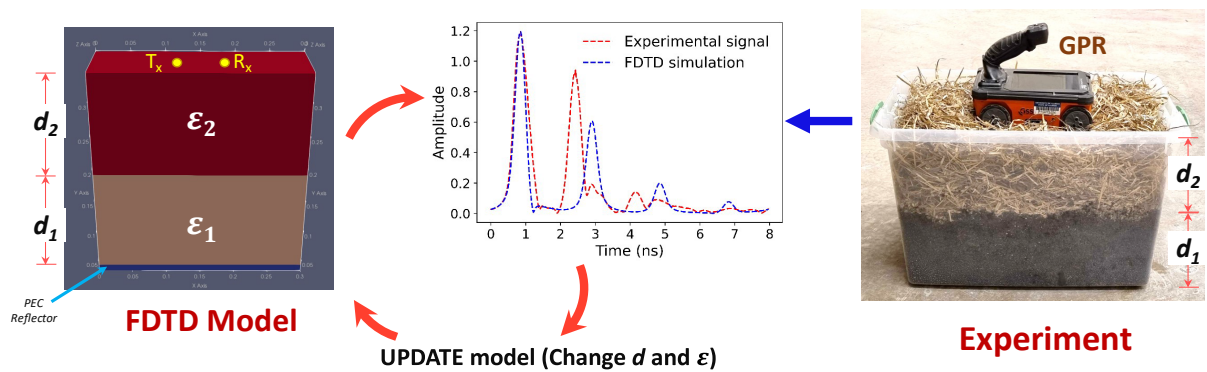

Figure 6: Parameter estimation through model updating.

### 2.3.1. Bayesian optimization

Initially, the proposed model-updating process was evaluated by estimating the expected values of the parameters of interest, such as dielectric permittivity, given the measured data (e.g., radar signals). Bayesian optimization is suitable for evaluating expensive objective functions and in our case, the evaluation of the function requires running expensive FDTD simulations at every iteration. Bayesian optimization operates by creating a Gaussian process-based posterior distribution of functions, aiming to capture the underlying nature of the objective function intended to optimize. As more observations are gathered, this posterior distribution refines itself, enhancing the algorithm's confidence in identifying which unexplored regions are to be further explored. In this paper, the objective function shown in Equation 3 is defined as the relative error (*RE*) between the experimentally measured signal ($y$) and the numerical simulation signal ($\bar{y}$).

### 2.3.2. Bayesian Model Updating (BMU) using MCMC

Probabilistic Bayesian model updating coupled with Markov Chain Monte Carlo (MCMC) simulation is an efficacious parameter estimation approach, offering several advantages over deterministic optimization methods. First, it allows for incorporating prior knowledge about the modeled system, which is particularly useful in cases with limited or noisy data. Prior knowledge can be obtained from past experience or data from experiments. Second, it provides a probabilistic framework for model calibration and uncertainty quantification. This means that not only can the model parameters be estimated, but the uncertainty associated with these estimates can also be quantified. This attribute is pivotal in many applications where the accuracy and reliability of the model are critical. Third, newly available data or evidence can be used to update the model parameters using Bayesian inference, leading to faster optimization of complex and high-dimensional problems.

The schematic representation of the Bayesian model updating (BMU) is shown in the Bayesian inference loop of Figure 3. The measured data ($y$) is obtained from the GPR experiment, and the experiment is replicated in the FDTD simulation. The FDTD model initialized with random parameters provides $\bar{y}$, which in turn enables the calculation of the error (Equation 4) and the likelihood function. The likelihood is then combined with the prior to provide the posterior distribution of the parameters using Bayes' theorem (Equation 5). In Equation 5, $P(\varepsilon|y)$ is the posterior distribution of the parameters, $P(y|\varepsilon)$ is the likelihood, $P(\varepsilon)$ is the prior and $P(y)$ is the normalizing term called the evidence. The distribution of dielectric permittivity and other parameters of a material layer is typically unknown before historical data collection programs are conducted. For this reason, a uniform/non-informative prior over the parameter space was assumed in our study. The normalizing term can be ignored, and the inferences about the model parameters can be drawn using Equation 6 (Qin et al., 2016).

$$\textbf{Error} = \sqrt{\frac{\sum(\boldsymbol{y} - \bar{\boldsymbol{y}})^2}{\sum \boldsymbol{y}^2}} \qquad \text{Equation 4}$$

$$\boldsymbol{P(\varepsilon|y)} = \frac{\boldsymbol{P(y|\varepsilon)P(\varepsilon)}}{\boldsymbol{P(y)}} \qquad \text{Equation 5}$$

$$\boldsymbol{P(\varepsilon|y)} \propto \boldsymbol{P(y|\varepsilon)P(\varepsilon)} \qquad \text{Equation 6}$$

The GPR data measured through experimental investigation can be described using Equation 7, where $f(\boldsymbol{\varepsilon})$ is the output from non-linear forward simulation and $error(\boldsymbol{\varepsilon})$ is the measurement error. To model the likelihood function, the measurement error has been assumed to be independent and normally distributed-$N(0,\sigma)$, with a mean of zero and a constant variance ($\sigma^2$). With this formulation, the likelihood function is defined as shown in Equation 8. Here, $n$ denotes the number of GPR data samples and $\sigma$ is the standard deviation of the measurement error). This function was converted to log-likelihood for computational convenience, and the log-likelihood function is shown in Equation 9. A better fit between numerical simulation and experimental measurement will result in smaller values of $(f_i(\boldsymbol{\varepsilon}) - \boldsymbol{y_i})^2$ and larger values of log-likelihood.

As mentioned before, the BMU problem is often solved by Markov Chain Monte Carlo (MCMC) simulation. In the MCMC sampling process, a set of criteria that rely solely on the posterior probability density function (PDF) and ignore the normalizing term, are used to decide whether to accept or reject the next sample. With the progress of the simulation in each step, the prior probability density is transformed, and it finally reaches the posterior distribution of each parameter.

$$y = f(\varepsilon) + error(\varepsilon) \qquad \text{Equation 7}$$

$$P(y|\varepsilon) = \frac{1}{(\sqrt{2\pi}\sigma)^n} \times \exp\left(-\frac{1}{2}\sum_{i=1}^{n}\left(\frac{f_i(\varepsilon) - y_i}{\sigma}\right)^2\right) \qquad \text{Equation 8}$$

$$P(y|\varepsilon) = \sum_{i=1}^{n}\left(-\frac{1}{2}\left(\frac{f_i(\varepsilon) - y_i}{\sigma}\right)^2 - \log\sqrt{2\pi}\sigma\right) \qquad \text{Equation 9}$$

## 2.4. Evaluation of estimated parameters

### 2.4.1. Time-domain reflectometry (TDR)

A TDR sensor was used in this study to measure the permittivity and corresponding moisture content of a soil sample (Topp et al., 1980; Wyseure et al., 1997) and compare them with values estimated by the proposed method. As shown in Figure 7, the sensor probe needs to be inserted into the soil, after which it provides a raw value based on the soil dielectric permittivity. The raw value is then converted to volumetric water content and relative permittivity by applying the calibration equations of the sensor. The TDR permittivity estimates were then compared to the permittivity values predicted by model updating to evaluate the correlation between them.

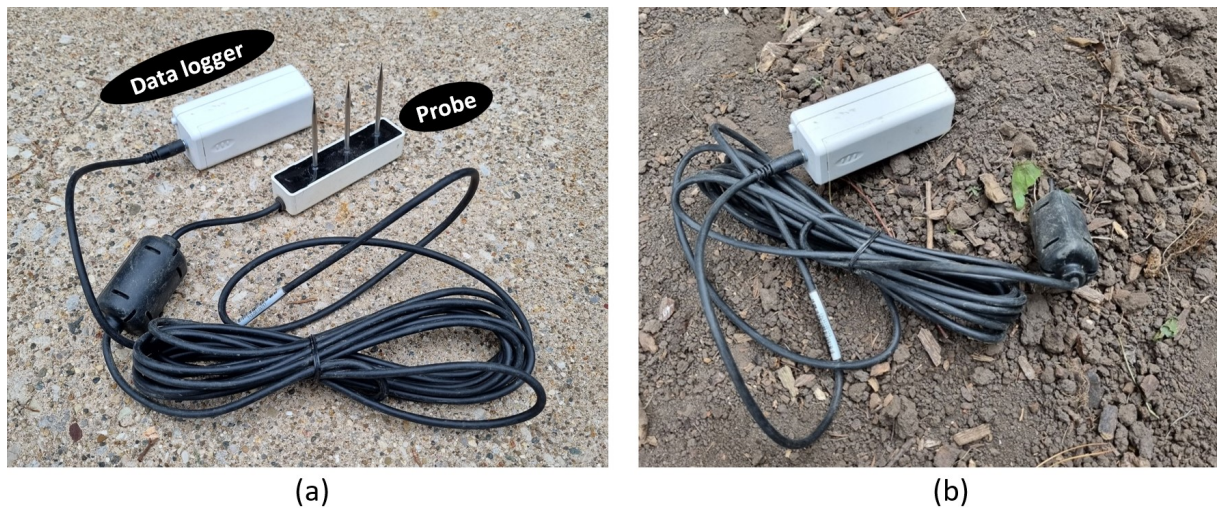

(a)          (b)

Figure 7: Measuring moisture content and permittivity by TDR.

### 2.4.2. Volumetric water content (VWC)

The moisture content of soil can be determined by gravimetric tests (by oven drying). In this process, oven drying of a soil sample is first carried out to determine the mass water content (MWC) of the soil, which is then multiplied by the bulk density of the soil to obtain the volumetric water content (VWC) (French et al., 1996). An estimation of the dielectric permittivity can then be obtained from VWC using Topp's equation shown in Equation 10 (Wu et al., 2022; Topp et al., 1980).

$$VWC = -0.053 + 0.0292\varepsilon_r - 5.5 \times 10^{-4}\varepsilon_r^2 + 4.3 \times 10^{-6}\varepsilon_r^3 \qquad \text{Equation 10}$$

where $\varepsilon_r$ denotes the dielectric permittivity.

Other dielectric models, such as the Dobson model (Dobson et al., 1985) and the Wang model (Wang & Schmugget, 1980), require additional input parameters like gravimetric clay and sand fraction, bulk density, and soil temperature (Zhang et al., 2023). These parameters may not be readily available, and their reliable measurements are challenging. Hence, Topp's model was used in this study for simplicity.

## 2.5. Experimental investigation

### 2.5.1. Laboratory investigations

The details of the experimental setup are depicted in Figure 8. The two material layers shown in the setup are a bottom soil layer of depth $d_1$, and an organic surface layer of depth $d_2$. The surface layer of vegetation and organic matter (or biomasss) on top of the soil layer is often called the '*top layer*' in this study. A steel reflector was placed underneath the soil layer, as shown in the schematic view. The estimation of soil permittivity and moisture in the presence of the top layer is analogous to measuring field soil moisture having a woody litter layer on top of the soil. To validate the proposed methodology for the estimation of soil permittivity for bare soil with no top layer as well as for soil covered with a top layer of organic material, a test matrix was formed by varying (i) top-layer depth, (ii) top-layer material type and (iii) soil-layer moisture content. The soil layer thickness ($d_1$) was kept at 15 cm throughout all the experiments, but the top layer depth ($d_2$) had four variations: 0 cm, 5 cm, 10 cm, and 15 cm. Three types of materials with three different coarseness and densities were used as the top layer. These are wood shavings, straw, and wood chips, with average particle sizes of 2.5 mm, 6.4 mm, and 11.7 mm, respectively. These experiments were repeated for an average of 7 different soil moisture contents ranging between 2-25 % VWC, and the top layer materials were used in their stock or natural condition for all experiments. GPR scans were recorded from the top of the material on the surface. Hence, the total number of GPR scans collected from the laboratory investigations was 84 (4 depths × 3 top layers × 7 soil moisture levels).

### 2.5.2. Field Investigations

The aim of the field investigation was to test whether the proposed method can estimate field soil moisture and track its change over time. The location of the measurement was on the east of the Newmark Civil Engineering Lab of the University of Illinois Urbana Champaign. A metal plate was buried at a depth of about 11 cm from the soil-top before starting data collection, and then GPR scans were collected from the top of bare soil for sixteen days.

Permittivity values of 16 days were predicted by the proposed method and then converted to volumetric moisture contents using the Topp's equation. The moisture contents were also measured with the TDR sensor at the same time as GPR scans, and soil samples were collected on day-14 and day-16 to determine moisture content by gravimetric tests. The composition and properties of the soil at the test location were not separately tested. However, as per the available soil report, the soil in the test area consists of about 24-25 % clay, 5-6 % sand, and 4-5 % organic matter.

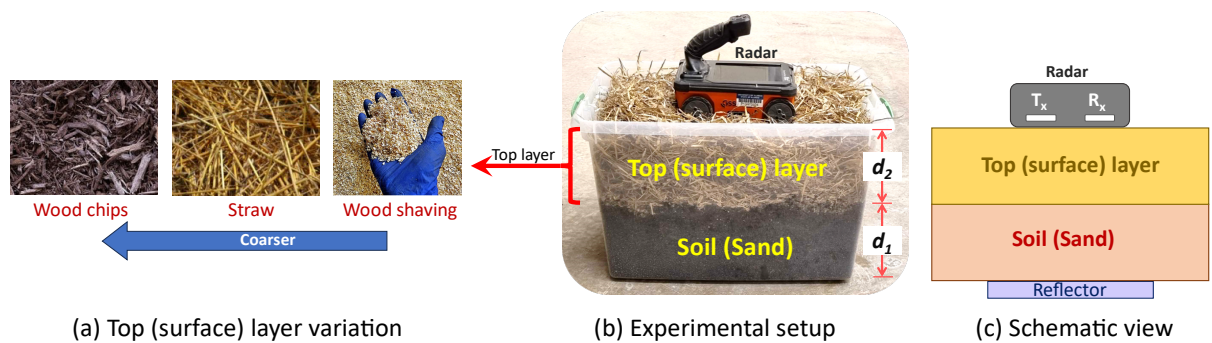

Figure 8: Details of the experimental setup.

## 3. Results

### 3.1. Laboratory investigations

#### 3.1.1. Estimation of permittivity

This section demonstrates the estimation of the permittivity of the soil layer using Bayesian optimization. The only model parameters estimated here are the dielectric permittivities of the soil and the top layer. Other parameters, including the depth of the two layers, were assumed to be known and kept constant in the model updating process.

The permittivity predicted by performing a GPR scan and optimization-based model updating is compared with the permittivity estimated by the TDR sensor, and the correlation between the two measurements is shown in Figure 9 for both soil-only and soil-plus top-layer cases. In every case, the lowest $R$ values are reported for the 15 cm soil + 15 cm top layer configuration. This is because there is higher attenuation due to the increased depth of the material, thus deteriorating the predictions and correlation for this configuration. The correlation also deteriorates as the average particle size of the top layer increases (wood chips have the largest average particle size), which is due to the attenuation and scattering of EM waves from coarser particles. These scenarios are also discussed in the 'Discussions' section (Section-4) using probability distributions.

**Correlation with analytical permittivity**

Analytical permittivity values were estimated by the travel-time method (Equation 1). To apply the travel-time method, the local maxima in the time-domain response were selected as peak amplitudes of the reflected waves (Zadhoush & Giannopoulos, 2022), and their corresponding times of arrivals were selected from the response. The permittivity values predicted by the proposed method were compared with the analytical values. Their correlation for two such cases is shown in Figure 10. The predicted values by model updating have a strong correlation with the analytical values. However, to accurately calculate analytical permittivity, choosing the correct peak and the time of arrival of that peak is essential. This is subject to subjective interpretation and requires an adequate understanding of the signal. In addition, when the peak from the bottom layer is not strong enough due to attenuation, accurate identification of its time of arrival becomes more challenging, if not impossible. A small deviation from the correct time of arrival can lead to a substantial error in permittivity estimation, as permittivity is proportional to the square of the time of flight. Besides, when multiple parameters such as permittivity and depth need to be estimated simultaneously, the analytical method will not suffice. These limitations can be overcome by the automated model updating-based parameter estimation proposed in this study.

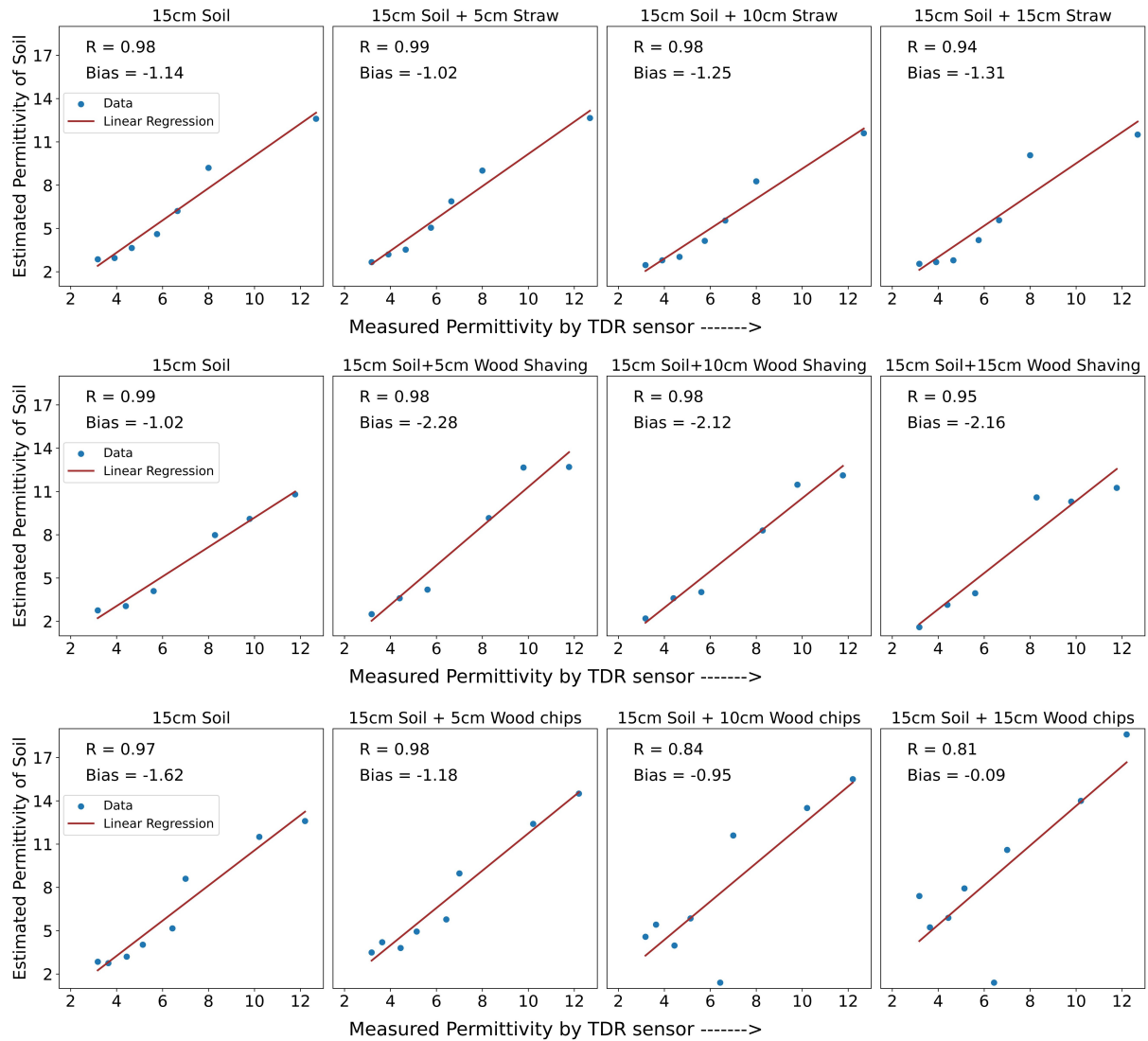

Figure 9: Correlation between estimated (predictions) and measured (TDR) permittivity with permittivity as the only model parameter.

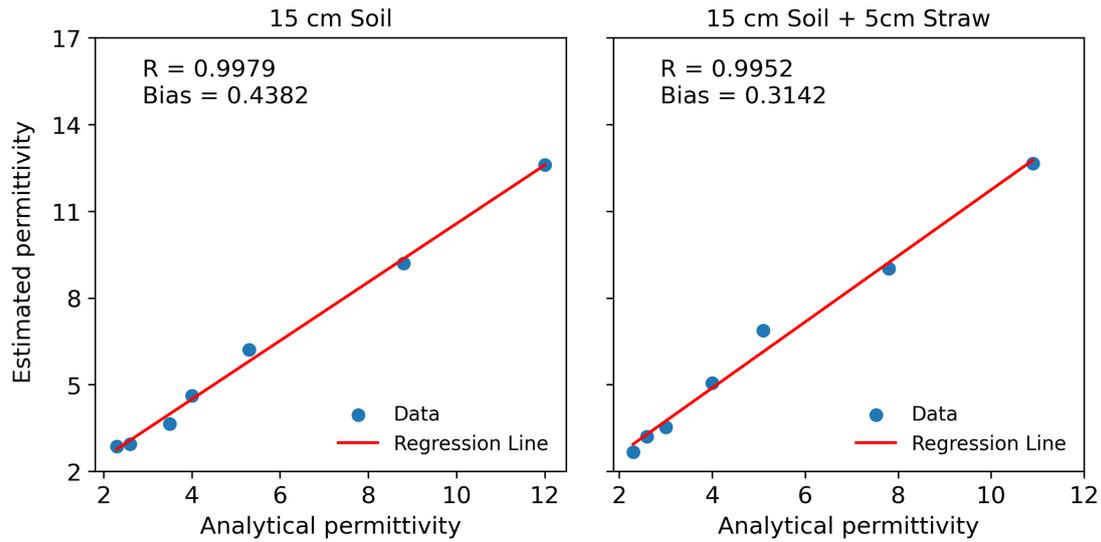

Figure 10: Correlation of estimated permittivity with analytical permittivity.

### 3.1.2. Estimation of permittivity and top (surface) layer depth

When soil moisture sensing is conducted using a reflector, the total depth from the radar at the top to the reflector at the bottom can usually be assumed to be known. However, the amount of material accumulated above the soil, i.e., the top layer depth, remains unknown due to its variation over time. Estimation of this top layer depth can also indicate where along the depth the soil layer begins. Estimating the soil permittivity and the layer depth simultaneously cannot be done by traditional analytical methods from a single GPR A-scan. However, the proposed methodology was employed while keeping multiple parameters as variables in the optimization algorithm. These parameters are the dielectric permittivity, electrical conductivity, and the top layer depth. After performing Bayesian optimization, the predicted permittivities of the soil layer in relation to the measured TDR permittivities are shown in Figure 11. This figure shows good correlations between the measured and estimated permittivities. However, the overall correlations decreased from the case when permittivity was the only variable. This is because the current case estimates five parameters instead of two in the previous case, which increases the dimensionality and complexity of the inverse problem. In addition, Figure 11 also shows that the predictions demonstrate a diminished correlation with an increase in top-layer material thickness. This can be associated with an increase in signal attenuation with the increase in material depth, thus resulting in weakened permittivity retrieval. Furthermore, the value of $R$ for (15 cm soil + 15 cm wood chips) was negative, which indicates poor correlation and predictions for this specific case. Anomalous predictions can be reasonably identified by leveraging probability distributions of the parameters that depict the uncertainty related to the estimates, as will be discussed in the 'Discussions' section (Section 4).

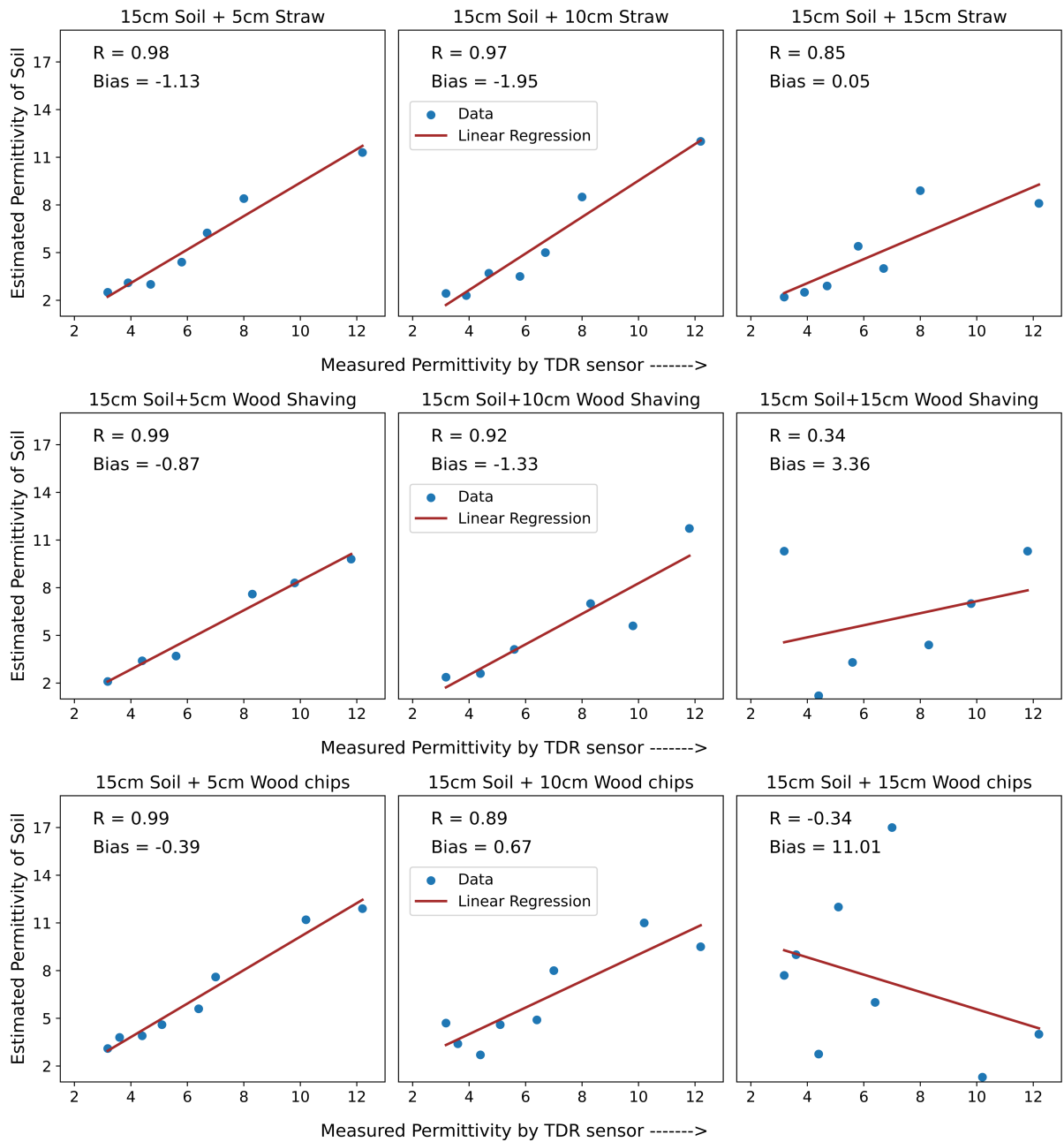

Figure 11: Correlation between estimated (predictions) and measured (TDR) soil permittivity when permittivity, conductivity, and depth are kept as variables.

The top layer depth was also predicted by the algorithm, and the summary of the predicted depths for all the experiments is shown with mean and standard deviation in Figure 12. The predicted depths are in reasonable agreement with the true depths, indicating the effectiveness of the method in determining top layer depth along with soil permittivity. The standard deviations for the cases of wood chips are higher for all depths. For 15 cm wood chips, the error and the standard deviation of predictions are both markedly high. This is because wood chips have larger particle sizes, which causes increased scattering and attenuation of the signal, thus resulting in weaker signal reconstruction and an increased likelihood of inconsistent predictions.

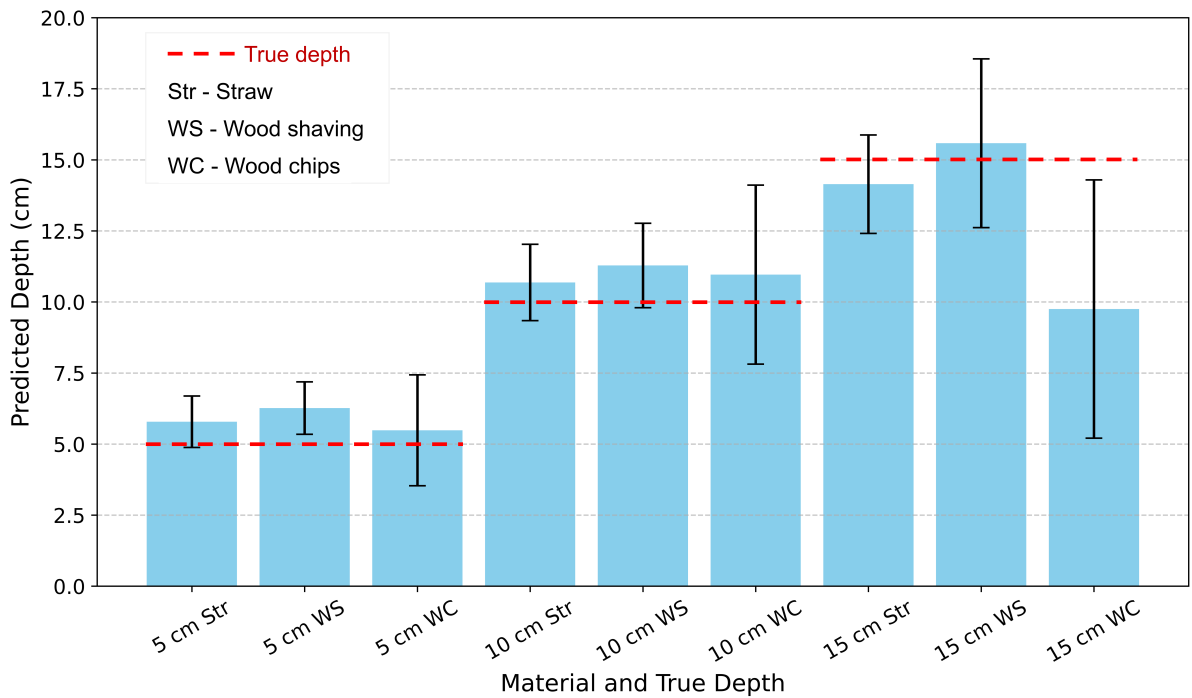

Figure 12: Predicted depth of the top layer material.

### 3.1.3. Probabilistic estimates

#### 3.1.3.1. Convergence of MCMC simulation

To provide an estimation of the uncertainty involved, the Markov Chain Monte Carlo (MCMC) technique was used to sample from the posterior distribution. Optimum values of the hyperparameters of the algorithm were used to balance computation time and the accuracy and convergence of the algorithm. MCMC diagnostics, including the trace plots, were checked to ensure the convergence of the chains, an example of which is depicted in Figure 13. In this figure, the progression of the chains over iterations (or step number) is shown in five plots corresponding to the five parameters estimated. Each plot here has seventeen chains shown with lines of different colors, illustrating how they explore the parameter space. After sufficient iterations/steps, the chains oscillate around a central value, indicating that they have reached a stationary distribution.

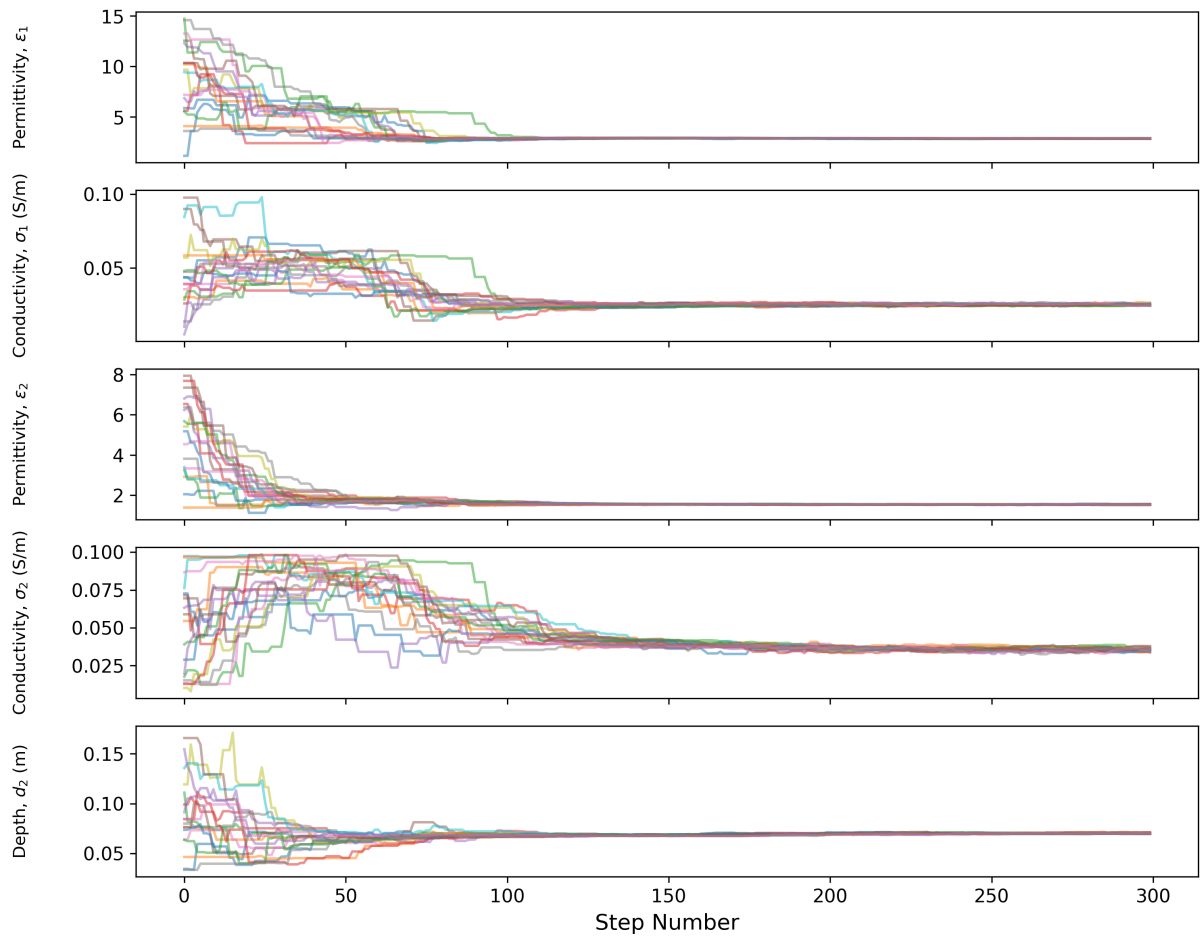

Figure 13: Trace plot showing the convergence of MCMC simulation.

### 3.1.3.2. Probability distributions of estimated parameters

The posterior distribution of the estimated parameters provides the probability distribution, indicating the uncertainty of the estimated values. The highest probabilities of the parameters were usually observed at the values that were predicted by the Bayesian optimization and presented in the previous section. As an example, the probabilistic estimates for one of the experiments (corresponding to 15 cm soil + 10 cm wood chips), along with their true/measured values, are shown in Figure 14.

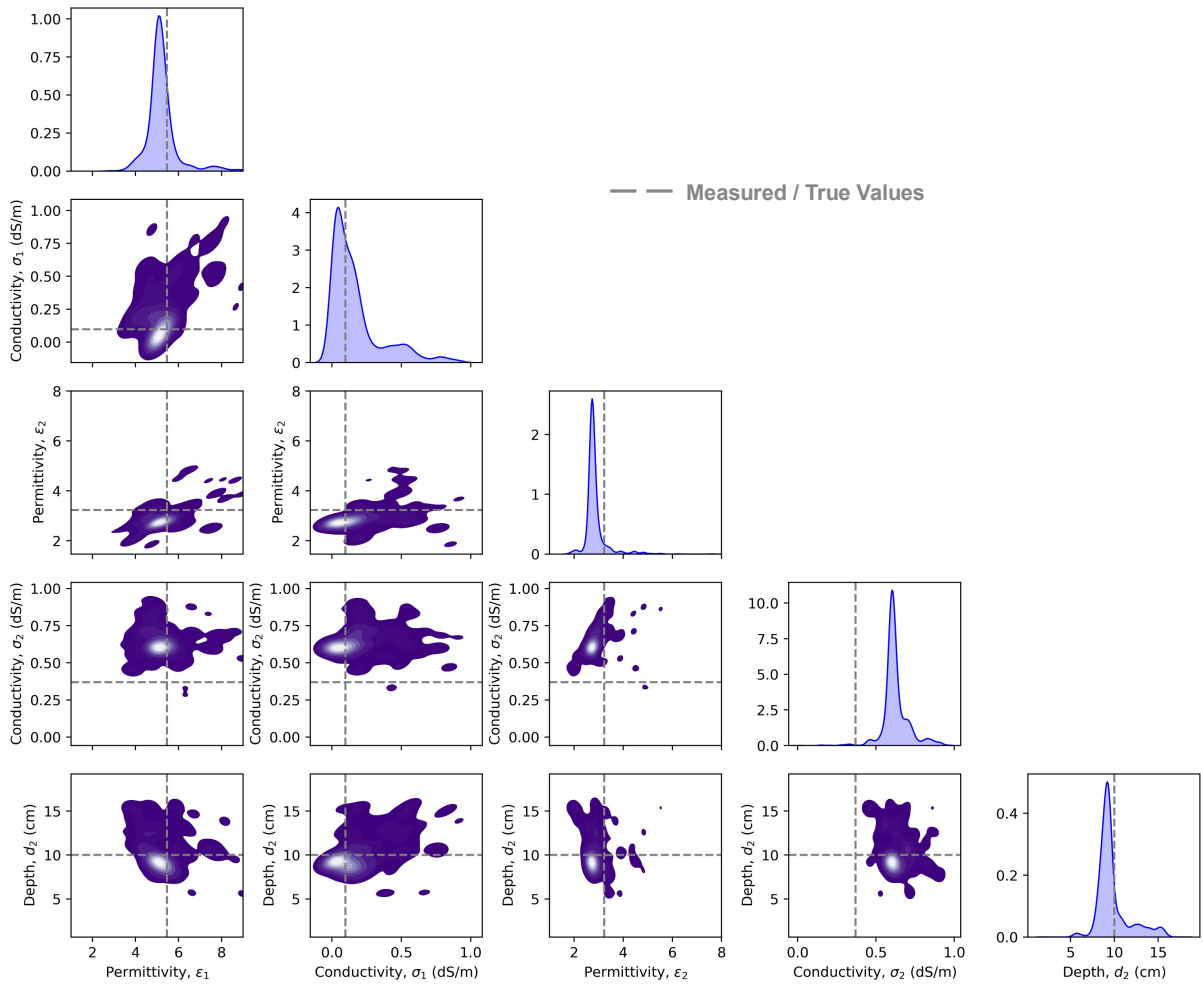

Figure 14: Individual and joint distribution of the estimated parameters.

Five parameters (permittivities and conductivities of the two layers and the depth of the top layer) were estimated by our proposed method, and in this figure, the diagonal plots show the distribution of the individual parameters, and the off-diagonal plot shows the joint distribution of different sets of two parameters. The true depth of the top layer ($d_2$) was 10cm, and the true soil permittivity ($\varepsilon_1$) was estimated to be 5.47. In the absence of direct measurement, this permittivity was found by measuring the water content via gravimetric tests and conversion to permittivity using Topp's equation. The electrical conductivity of both layers and the permittivity of the top layer were estimated by the TDR sensors. From the individual distributions in Figure 14, the maximum-a-posteriori (MAP) value or the mode of the posterior distribution was observed to be near the true or estimated values. The lighter colors in the joint distributions indicate values of high probability density, and those are mostly observed to be in the vicinity of the true/estimated values. The deviation of the conductivity measurement from the predicted values can be attributed to the limitation of the TDR sensor in measuring the conductivity of coarse woody materials.

## 3.2. Field investigations

For field study (Figure 15a), the comparison of the predicted moisture contents with TDR measurements and gravimetric tests is shown in Figure 15b. The approximate rainfall periods are also shown here, after each of these periods, depicting a steep rise in the moisture content of the soil. The overall trend of the predicted values matches the trend of the TDR measurements, and the moisture contents estimated by the gravimetric tests are also reasonably close to the predicted values. The slight deviation from the gravimetric results can be attributed to the fact that the bulk density of the soil in field conditions was not precisely the same as that in the laboratory test. From these results, it can be concluded that the proposed method can be applied to field moisture estimation.

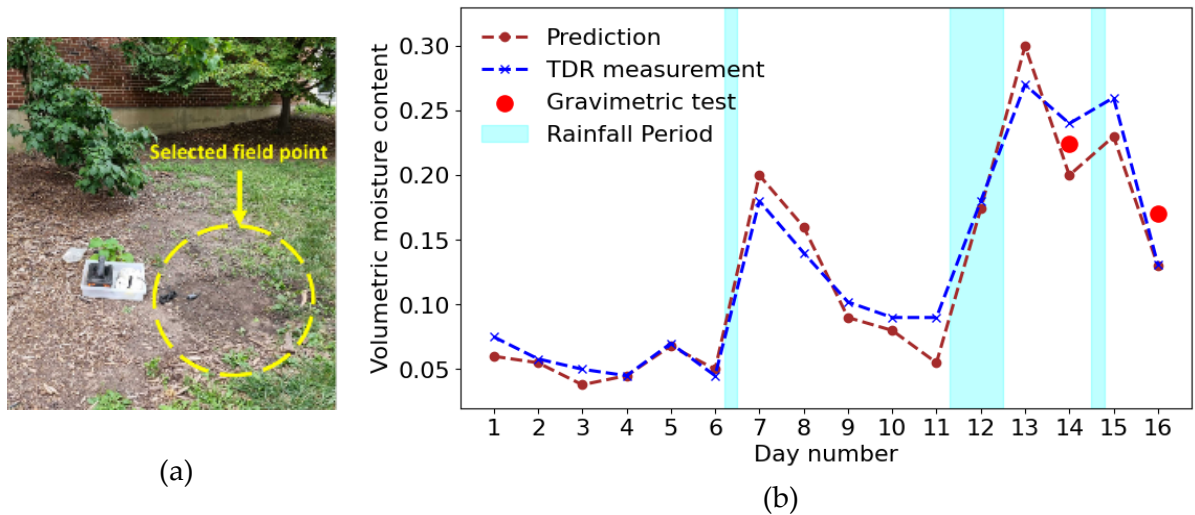

Figure 15: (a) location of the field test, (b) volumetric soil moisture obtained by the proposed method, TDR, and gravimetric tests.

## 4. Discussions

### 4.1. Performance on investigated scenarios

The reported results demonstrated the overall success of the proposed technique in the estimation of the target variables. A closer look at the estimations, their correlation with in situ measurements, and the resulting uncertainty estimations sheds light on the range of applications of the proposed technique. The proposed method in this study reports an $R$-value (between predicted soil permittivity and TDR permittivity) of about 0.84-0.99 with the presence of up to 10 cm of overlaying organic top layer. In recent studies reported in the literature, the correlation ($R$-value) between predicted soil moisture and other comparable parameters, such as in-situ measurements or radar backscatter, has been reported to lie within a range of 0.69 to 0.95 (Pathirana et al., 2024; Fluhrer et al., 2024; Sharma et al., 2018). However, the correlation in our work deteriorates for a higher depth of overlaying layer due to signal attenuation. As an illustration, an example from the case of 15 cm soil covered with 15cm wood chips is demonstrated in Figure 16. In this case, the minimum objective function (relative error obtained using Equation 3) results from a top layer depth equal to 2.8 cm, which is significantly smaller than the true depth of 15 cm. Figure 16 shows a comparison between the signal match using the true depth of 15cm (Figure 16b), with that using 2.8 cm (Figure 16a). It shows that the reflection from the bottom is not significant enough to influence the objective function of the optimization. This resulted in the value of the objective function being comparable to that from a different layer configuration, hence making the detectability of the true optimal solution impossible. Such high attenuation and reduction in signal strength are observed as both the permittivity and the total depth increase. In these scenarios, using a lower-frequency GPR may produce stronger signal amplitude from deeper layers and should be the subject of future studies. This is in line with the widely recognized trade-off between the frequency of GPR and its penetration depth, which should be considered in determining the range of applicability of permittivity retrieval. Nevertheless, the results of the experiments at lower depths clearly demonstrate the validity of the technique for organic top layer depths up to 10 cm.

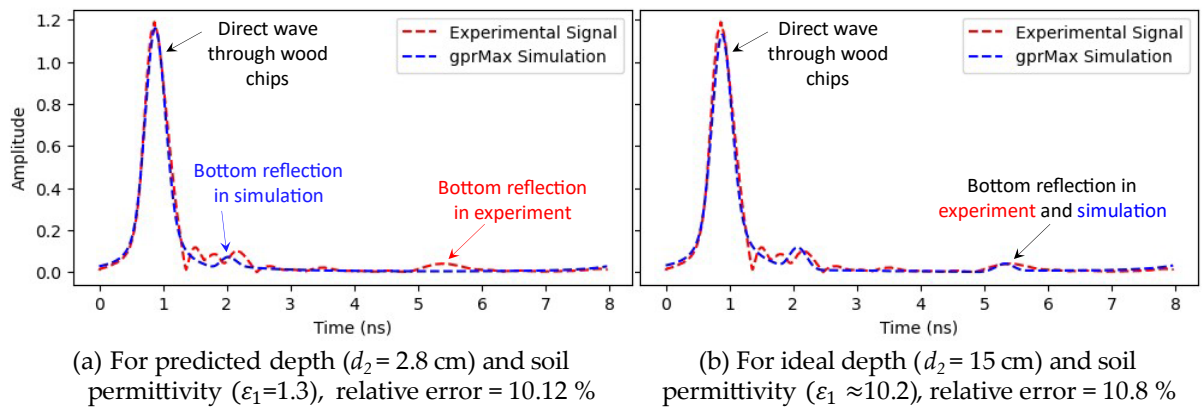

(a) For predicted depth ($d_2$ = 2.8 cm) and soil permittivity ($\varepsilon_1$=1.3), relative error = 10.12 %

(b) For ideal depth ($d_2$ = 15 cm) and soil permittivity ($\varepsilon_1 \approx$ 10.2), relative error = 10.8 %

Figure 16: Overlaid plots for (15 cm soil + 15 cm wood chips) corresponding to TDR soil permittivity of 10.2.

## 4.2. Uncertainty and identification of inconsistent predictions

An advantage of the proposed method over other deterministic inversion techniques is that it provides probabilistic estimates and quantifies the uncertainty of predicted soil moisture and other parameters. The obtained probabilistic estimates were further inspected to investigate how the top layer configuration affects the prediction in terms of probabilistic confidence and uncertainty. The probability distributions of the predicted soil permittivity are shown in Figure 17 for two variables, viz., depth of the top layer (Figure 17a) and particle size of the top layer material (Figure 17b). As the depth of the top layer decreases, the probability distribution curve shows a higher probability density at the expected value (TDR measured) and a narrower spread, indicating lower standard deviation and higher confidence in the prediction. This can be explained by the fact that as the layer depth decreases, there is lower attenuation, and the reflection from the reflector is stronger, rendering more information in the signal received by the GPR. Similarly, for higher values of top-layer depth, the signal carries less information due to higher attenuation, which leads to a prediction with lower confidence. This attenuation of signal due to increasing layer depth was also evident in our initial analysis of signal variations due to changes in material parameters (Figure 2a).

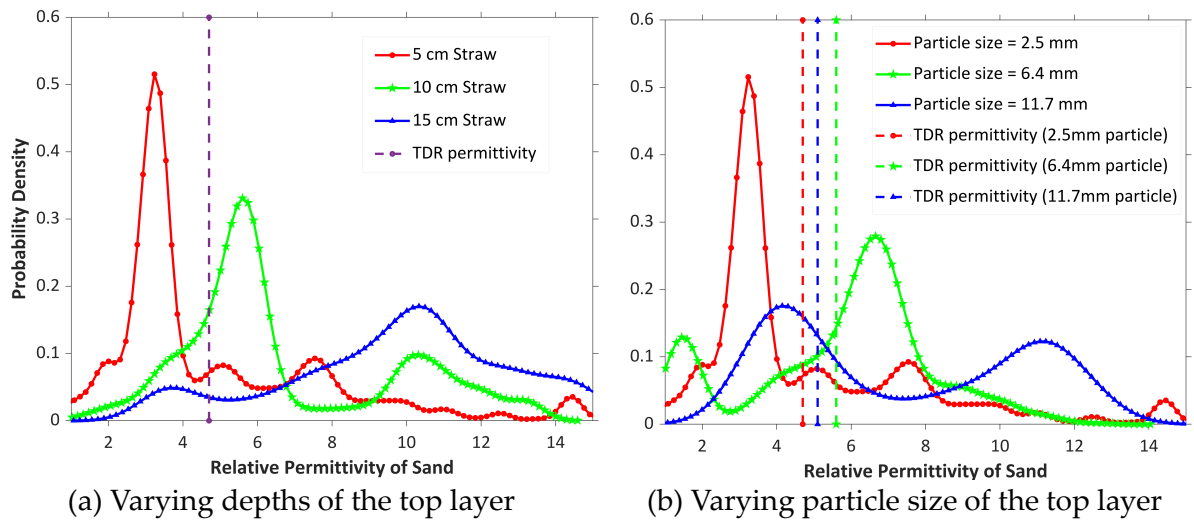

(a) Varying depths of the top layer   (b) Varying particle size of the top layer

Figure 17: Distributions of estimated permittivity for varying top layer depth and particle size.

A similar phenomenon can be observed in the case of varying particle sizes of the top layer. As the particle size increases, there is more noise and scattering from the coarser particles and, thus, higher attenuation in the received signal. This leads to predictions with lower confidence levels and higher standard deviations. It can also be observed that the peak value for 15 cm straw depth in Figure 17(a) and the second peak (at permittivity≅11.5) for 11.7 mm particle size in Figure 17(b) significantly differ from the TDR measurements. The second peak in Figure 17(b) represents the value that can produce a close fit between the numerical and the experimental signal, similar to the case shown in Figure 16, where wrong depth prediction yielded a lower mismatch between the two signals. In these cases, the wider distribution and lower probability density achieved by the Bayesian inference indicate higher uncertainty. Thus, lower confidence and higher uncertainty in prediction obtained by the proposed method are indicators that can facilitate the identification of relatively inconsistent predictions similar to the ones mentioned above.

### 4.3.  Advantages over other techniques

In addition to the probabilistic predictions, the proposed radar sensing method presents other desirable advantages over existing techniques. Foremost, it obviates the need to use battery-powered devices and labor-intensive gravimetric tests for soil moisture prediction. The permittivity values predicted with this method correlate well with TDR permittivity, analytical permittivity, and the moisture content estimated by conventional gravimetric tests.

In addition to the soil moisture content, the proposed method can also simultaneously estimate the depth of the overlaying layer, which is crucial for wildfire risk assessment.

Furthermore, the method does not require many repeated measurements at the same point to obtain probability distributions. Instead, the Bayesian inference leverages the power of the prior and likelihood to update belief and obtain stable probability distributions of the parameter values based on the discrepancy between the observed experimental data and numerical responses from model updating.

### 4.4. Range of applicability to other problems

The proposed method can be extended beyond the estimation of soil moisture and vegetation layer depth to various other applications. FWI has been used for estimating moisture content in building materials (Klewe et al., 2021), corrosion monitoring in concrete structures (Kalogeropoulos et al., 2011), reconstruction of tunnel lining defects (Feng et al., 2019), monitoring cement setting time (Xie et al., 2022), inspecting tree trunk defects (Feng et al., 2023), hydraulic aquifer characterization (Haruzi et al., 2022), to name a few. These applications often employ point radar measurements similar to the current study but primarily with a deterministic approach without information on uncertainties. Hence, the method proposed in this study can potentially be used as an uncertainty-aware approach for these applications.

### 4.5. Limitations and future works

While the proposed technique shows great promise in the measurement of soil and surface layers, there are a few limitations and areas of improvement that can be the subject of future studies. As discussed before, the weakened correlation due to the attenuation resulting from excessive depth can be associated with the penetration capability or the frequency of the GPR antenna. The discussion to overcome this limitation leads to the well-known tradeoff between resolution and penetration of EM signals as a function of frequency. To harness the strengths of both high and low frequencies, future studies can focus on the simultaneous inclusion of multiple frequencies. This can also help remedy the problem of non-uniqueness in optimization. Further detailed investigation on other aspects, such as random woody particle size distribution and porosity of soil and the top surface layer and their effect on wave propagation and attenuation, can be the subject of future studies. In addition, while the current study focuses on time-domain signals only, future studies can combine time-domain signal, frequency, and power spectra for more robust parameter estimation. The feasibility and robustness of the proposed parameter estimation for multiple layers can also enable material characterization in more complex applications, such as understory ground layer estimations from radars mounted on above-canopy Uncrewed Aerial Systems (UAS). The current method can also be extended for application to data collected by a moving radar like that mounted on a UAS.

The present study uses a metal plate to define the soil bottom, similar to previous studies (Ding et al., 2023; Anbazhagan et al., 2020; Cui et al., 2012; Lambot et al., 2004). Although it obviates the need to bury and maintain battery-powered devices inside the soil (Ding et al., 2023), burying metal plates adds a degree of cost and labor. In addition, the measurements with the metal plate are point measurements. Hence, future studies can focus on expanding the scale of measurements and the potential of applying the proposed model-updating approach without the metal reflector.

For parameter estimation, alternative methods such as supervised machine learning models (e.g., SVM Regression, Gradient Boosting, and Random Forest Regression) can also be used to correlate the input radar signals with the target parameters, but they require labeled data for training the models. With the availability of an adequate amount of training data, these methods can be used in future investigations.

## 5. Conclusions

This paper proposes a Bayesian inversion-based model-updating methodology to estimate sub-surface properties, including depth, dielectric permittivity, and electrical conductivity of layers of soil and organic overlaying material. Model updating was performed through FDTD simulations, and the methodology was validated on experimental data from laboratory and field investigations. The parameters of the transmitted radar pulse in FDTD simulation were first optimized and calibrated, after which FDTD simulations could closely replicate real experimental signals collected by a commercial GPR. Laboratory investigations were carried out to predict permittivities of both bare soil and soil with an overlaying organic material layer having varying levels of coarseness. Field soil moisture was also predicted for sixteen days using the proposed method. The predictions were correlated and compared with measurements obtained by TDR and gravimetric tests. The following conclusions can be drawn from the investigations:

- The proposed method can simultaneously estimate multiple parameters of material layers from GPR signals, including depth, permittivity, and conductivity, while the traditional analytical method can only estimate a single parameter (e.g., permittivity), assuming other parameters are known (e.g., depth).
- The proposed method can provide probabilistic estimates of sub-surface permittivity, moisture content, conductivity, and depth using Bayesian inference with the MCMC algorithm. The probabilistic predictions provide the distribution of the estimated parameters and the uncertainty related to those predictions without the need for repeated experimental measurements.

- The predictions in the laboratory investigations are in good agreement with TDR measurements for bare soil and soil with an overlaying material layer of up to 10 cm depth. Predicted soil permittivity or water content is also in good agreement with values estimated by gravimetric tests.
- For higher depth (> 10 cm) and coarser particle size of the top (surface) layer material, the attenuation of the received GPR signal results in inconsistent permittivity predictions and diminished correlation with TDR measurements. However, these inconsistent predictions showed lower probability density and larger standard deviation, indicating higher uncertainty compared to the accurate predictions. As such, uncertainty from the proposed estimation method can be leveraged to identify such inconsistent predictions.
- The predicted field soil moisture content is also in good agreement with the trend of the TDR measurements and moisture content measured by gravimetric tests, validating the effectiveness of the proposed method for field moisture estimation.

Overall, it was shown that the proposed method can estimate sub-surface depth, dielectric permittivity, and moisture content, which can be used for a wide range of applications spanning precision agriculture, environmental sensing, and moisture mapping to wildfire risk assessment and non-destructive evaluation (NDE) in engineering structures.

## Acknowledgments

This study was in part funded by the United States Forest Service and Keysight Technologies. The authors sincerely appreciate this support. The authors would also like to express their sincere appreciation to the NVIDIA Corporation for their generous support with the donation of GPUs used in this study. The authors also sincerely thank Yuxiang Zhao and Kurt Soncco for assistance with computing, Anshu Abhinav for assistance with TDR operation, and Coleman Froehlke for assistance during data collection and gravimetric tests. The authors further thank Matthew Ware and the GSSI technical team for their technical support.

## Data and code availability

The Python code used to generate the findings of this study is available in the following GitHub repository: https://github.com/ishfaqa2/Code-Bayesian-Inversion-of-GPR-waveforms-. A sample of the experimental data is also available in the same repository.

# List of Figures